\documentclass[12pt]{article}

\usepackage[letterpaper, margin=1in]{geometry}

\usepackage[colorlinks]{hyperref}
\usepackage{makeidx}
\usepackage{framed}
\usepackage{fancyvrb}
\usepackage{color}
\usepackage{hyperref}
\hypersetup{colorlinks = true, allcolors = blue}
\usepackage{float}
\usepackage{flafter}
\usepackage{tabularx}
\usepackage{booktabs}
\usepackage{longtable}
\usepackage{multirow}
\usepackage{graphicx}
\usepackage{afterpage}

\usepackage{datetime}

\yyyymmdddate

  \usepackage{setspace}
  \AtBeginEnvironment{tabular}{\singlespacing}
  \AtBeginEnvironment{table}{\singlespacing}
  \AtBeginEnvironment{longtable}{\singlespacing}
  \AtBeginEnvironment{verbatim}{\singlespacing}
  \AtBeginEnvironment{Shaded}{\singlespacing}

\NewDocumentCommand\citeproctext{}{}

\makeatletter
 \let\@cite@ofmt\@firstofone
 \def\@biblabel#1{}
 \def\@cite#1#2{{#1\if@tempswa , #2\fi}}
\makeatother
\newlength{\cslhangindent}
\newlength{\csllabelwidth}
\newenvironment{CSLReferences}[2] 
 {\begin{list}{}{%
  \setlength{\itemindent}{0pt}
  \setlength{\leftmargin}{0pt}
  \setlength{\parsep}{0pt}
  \ifodd #1
   \setlength{\leftmargin}{\cslhangindent}
   \setlength{\itemindent}{-1\cslhangindent}
  \fi
  \setlength{\itemsep}{#2\baselineskip}}}
 {\end{list}}
\usepackage{calc}

  \usepackage{amsmath}
  \usepackage{amssymb}

\usepackage{xcolor}
\definecolor{dark}{HTML}{2c2e35}
\color{dark}

\definecolor{myblue}{HTML}{1e3765}
\hypersetup{colorlinks,linkcolor=myblue,urlcolor=myblue}

\makeindex

  \usepackage{color}
  \usepackage{fancyvrb}

  \DefineVerbatimEnvironment{Highlighting}{Verbatim}{commandchars=\\\{\}}
  \usepackage{framed}
  \definecolor{shadecolor}{RGB}{241,243,245}
  \newenvironment{Shaded}{\begin{snugshade}}{\end{snugshade}}

  \newcommand{\AttributeTok}[1]{\textcolor[rgb]{0.40,0.45,0.13}{#1}}

  \newcommand{\CommentTok}[1]{\textcolor[rgb]{0.37,0.37,0.37}{#1}}
  
  \newcommand{\ConstantTok}[1]{\textcolor[rgb]{0.56,0.35,0.01}{#1}}
  \newcommand{\ControlFlowTok}[1]{\textcolor[rgb]{0.00,0.23,0.31}{\textbf{#1}}}
  
  \newcommand{\DecValTok}[1]{\textcolor[rgb]{0.68,0.00,0.00}{#1}}

  \newcommand{\FloatTok}[1]{\textcolor[rgb]{0.68,0.00,0.00}{#1}}
  \newcommand{\FunctionTok}[1]{\textcolor[rgb]{0.28,0.35,0.67}{#1}}

  \newcommand{\NormalTok}[1]{\textcolor[rgb]{0.00,0.23,0.31}{#1}}
  
  \newcommand{\OtherTok}[1]{\textcolor[rgb]{0.00,0.23,0.31}{#1}}

  \newcommand{\SpecialCharTok}[1]{\textcolor[rgb]{0.37,0.37,0.37}{#1}}
  
  \newcommand{\StringTok}[1]{\textcolor[rgb]{0.13,0.47,0.30}{#1}}

\author{
  Mauricio Vargas Sepúlveda (ORCID 0000-0003-1017-7574)\\Department of
Political Science, University of Toronto\\Munk School of Global Affairs
and Public Policy, University of Toronto\\
  \smallskip\\
  \\\\
  \smallskip\\
  Corresponding author: m.sepulveda@mail.utoronto.ca
}
\title{Replicating The Log of Gravity}
\date{Last updated: \today\ \currenttime}

\begin{document}

\maketitle

\thispagestyle{empty}
\tableofcontents
\setcounter{page}{0}
\clearpage

\afterpage{\setlength\parskip{10pt}}

\section{Abstract}\label{abstract}

This document replicates the main results from Santos Silva and Tenreyro
(2006) in R. The original results were obtained in TSP back in 2006. The
idea here is to be explicit regarding the conceptual approach to
regression in R. For most of the replication I used base R without
external libraries except when it was absolutely necessary. The findings
are consistent with the original article and reveal that the replication
effort is minimal, without the need to contact the authors for
clarifications or incur into data transformations or filtering not
mentioned in the article.

\section{Introduction}\label{introduction}

Santos Silva and Tenreyro (2006) used
\href{https://en.wikipedia.org/wiki/TSP_(econometrics_software)}{TSP}
back in 2006 to estimate different functional forms of the gravity model
of trade. The article is a classic in the field of international trade
and has been cited over 8,500 times according to Google Scholar. Its
contribution, which was a new Poisson-based estimator to correct the
bias in the Ordinary Least Squares (OLS) estimator under
heteroskedasticity, has been widely used in the academic literature.
Besides pure academic interests, the OLS estimator bias has implications
for policy analysis, and posterior works with clear insights for
International Relations academics and Public Policy practitioners,
including Head and Mayer (2014), Yotov et al. (2016), and Felbermayr et
al. (2020).

Outside of statistical concerns about biases, the field of International
Relations has developed interesting critiques about leaving theory
behind. Mearsheimer and Walt (2013) argues that the field has been
dominated by quantitative studies that present simplistic hypothesis
after incentives to present academic findings as more interesting for
policymakers. From this critique, Santos Silva and Tenreyro (2006) is
relevant not only for its methodological contribution and presenting an
estimator that can largely differ from OLS, but also for presenting an
estimator that is consistent with micro-founded approaches to the
gravity model of trade, including Eaton and Kortum (2001).

The goal in this document is to be explicit regarding the conceptual
approach to regression in R. For most of the replication we used base R
(R Core Team 2021) and relying on external libraries (e.g., R packages)
when it was absolutely necessary based on the criteria that implementing
methods such as the Tobit estimator from scratch would be time consuming
and would deviate from the replication effort. The results are
consistent with the gravity package (Woelwer et al. 2020) which provides
convenient wrappers for gravity estimation, but the idea was to be
explicit about the steps involved in the replication process and using
wrappers would have made the steps less clear.

\section{Original codes and data}\label{original-codes-and-data}

We organized the original codes and data from the authors' site,
\href{https://personal.lse.ac.uk/tenreyro/lgw.html}{The Log of Gravity},
on \href{https://github.com/pachadotdev/log-of-gravity}{GitHub} and
therefore ease replication in the event that the links change in the
future.

We obtained the original data and code with the following R code:

\begin{Shaded}
\begin{Highlighting}[]
\NormalTok{url }\OtherTok{\textless{}{-}} \StringTok{"https://personal.lse.ac.uk/tenreyro/regressors.zip"}
\NormalTok{zip }\OtherTok{\textless{}{-}} \FunctionTok{gsub}\NormalTok{(}\StringTok{".*/"}\NormalTok{, }\StringTok{""}\NormalTok{, url)}
\ControlFlowTok{if}\NormalTok{ (}\SpecialCharTok{!}\FunctionTok{file.exists}\NormalTok{(zip)) }\FunctionTok{try}\NormalTok{(}\FunctionTok{download.file}\NormalTok{(url, zip))}

\NormalTok{dout }\OtherTok{\textless{}{-}} \StringTok{"regressors"}
\ControlFlowTok{if}\NormalTok{ (}\SpecialCharTok{!}\FunctionTok{dir.exists}\NormalTok{(dout)) }\FunctionTok{unzip}\NormalTok{(zip, }\AttributeTok{exdir =}\NormalTok{ dout)}
\end{Highlighting}
\end{Shaded}

The original dataset obtained in the previous step is available in Stata
format, a closed source format that we can read in R thanks to the haven
package (Wickham and Miller 2021) without loss of information or other
common problems when reading proprietary formats.

We proceeded to import R packages to ease our work and read the data
with the following R code:

\begin{Shaded}
\begin{Highlighting}[]
\FunctionTok{library}\NormalTok{(haven) }\CommentTok{\# read Stata datasets}
\FunctionTok{library}\NormalTok{(censReg) }\CommentTok{\# Tobit estimation}
\FunctionTok{library}\NormalTok{(stargazer) }\CommentTok{\# table of results}

\NormalTok{log\_of\_gravity }\OtherTok{\textless{}{-}} \FunctionTok{read\_dta}\NormalTok{(}\FunctionTok{paste0}\NormalTok{(dout, }\StringTok{"/Log of Gravity.dta"}\NormalTok{))}
\end{Highlighting}
\end{Shaded}

\section{Models replication}\label{models-replication}

\subsection{Poisson Pseudo Maximum
Likelihood}\label{poisson-pseudo-maximum-likelihood}

Table 3 in Santos Silva and Tenreyro (2006) summarises a large portion
of the article. Starting from the Poisson Pseudo Maximum Likelihood
(PPML) estimation, we can replicate the results with the Generalized
Linear Model (GLM) and a quasi-Poisson link in R. The original article
estimates the model with and without zero flows.

We could replicate the PPML model results with and without zero flows
with the following R code:

\begin{Shaded}
\begin{Highlighting}[]
\NormalTok{ppml\_formula }\OtherTok{\textless{}{-}}\NormalTok{ trade }\SpecialCharTok{\textasciitilde{}}\NormalTok{ lypex }\SpecialCharTok{+}\NormalTok{ lypim }\SpecialCharTok{+}\NormalTok{ lyex }\SpecialCharTok{+}\NormalTok{ lyim }\SpecialCharTok{+}\NormalTok{ ldist }\SpecialCharTok{+}\NormalTok{ border }\SpecialCharTok{+}
\NormalTok{  comlang }\SpecialCharTok{+}\NormalTok{ colony }\SpecialCharTok{+}\NormalTok{ landl\_ex }\SpecialCharTok{+}\NormalTok{ landl\_im }\SpecialCharTok{+}\NormalTok{ lremot\_ex }\SpecialCharTok{+}\NormalTok{ lremot\_im }\SpecialCharTok{+}
\NormalTok{  comfrt\_wto }\SpecialCharTok{+}\NormalTok{ open\_wto}

\NormalTok{fit\_ppml\_1 }\OtherTok{\textless{}{-}} \FunctionTok{glm}\NormalTok{(}
\NormalTok{  ppml\_formula,}
  \AttributeTok{data =}\NormalTok{ log\_of\_gravity,}
  \AttributeTok{subset =}\NormalTok{ trade }\SpecialCharTok{\textgreater{}} \DecValTok{0}\NormalTok{,}
  \AttributeTok{family =} \FunctionTok{quasipoisson}\NormalTok{()}
\NormalTok{)}

\NormalTok{fit\_ppml\_2 }\OtherTok{\textless{}{-}} \FunctionTok{glm}\NormalTok{(}
\NormalTok{  ppml\_formula,}
  \AttributeTok{data =}\NormalTok{ log\_of\_gravity,}
  \AttributeTok{family =} \FunctionTok{quasipoisson}\NormalTok{()}
\NormalTok{)}
\end{Highlighting}
\end{Shaded}

The replication effort for the PPML model was minimal, it was sufficient
to look at the summary table in the article and subset the data to drop
zero flows. This step did not require to guess data cleaning steps or
transformations not mentioned in the article.

\subsection{Ordinary Least Squares}\label{ordinary-least-squares}

The only consideration for the OLS model was to drop zero flows for some
of the models with log in the dependent variable even when Table 3 is
not explicit about this, otherwise we break the fitting algorithm
(e.g.~because \(\log(0) \to -\infty\)).

For estimations of the form
\(\log(\text{trade}) = \beta_0 + \beta_1 \text{lypex} + \dots + \varepsilon,\)
we needed to drop zero flows to replicate the result. On the other hand,
for estimations of the type
\(\log(1 + \text{trade}) = \beta_0 + \beta_1 \text{lypex} + \dots + \varepsilon,\)
we did not need to drop zero flows.

\begin{Shaded}
\begin{Highlighting}[]
\NormalTok{fit\_ols\_1 }\OtherTok{\textless{}{-}} \FunctionTok{lm}\NormalTok{(}
  \FunctionTok{update.formula}\NormalTok{(ppml\_formula, }\FunctionTok{log}\NormalTok{(.) }\SpecialCharTok{\textasciitilde{}}\NormalTok{ .),}
  \AttributeTok{data =}\NormalTok{ log\_of\_gravity,}
  \AttributeTok{subset =}\NormalTok{ trade }\SpecialCharTok{\textgreater{}} \DecValTok{0}
\NormalTok{)}

\NormalTok{fit\_ols\_2 }\OtherTok{\textless{}{-}} \FunctionTok{lm}\NormalTok{(}
  \FunctionTok{update.formula}\NormalTok{(ppml\_formula, }\FunctionTok{log}\NormalTok{(}\DecValTok{1} \SpecialCharTok{+}\NormalTok{ .) }\SpecialCharTok{\textasciitilde{}}\NormalTok{ .),}
  \AttributeTok{data =}\NormalTok{ log\_of\_gravity}
\NormalTok{)}
\end{Highlighting}
\end{Shaded}

\subsection{Tobit}\label{tobit}

The Tobit estimation required the use of the censReg package (Henningsen
2020). Unlike PPML and OLS models, this required us to extract the right
hand side of the model formula to define a vector of zeroes with a
length equal to the independent variables plus two as starting point for
the Maximum Likelihood estimation (e.g., counting the depending variable
and intercept besides the estimating slopes).

In order to obtain the \(a\) parameter that matches the results in the
article we proceeded with an iteration loop until achieving convergence
with respect to one of the estimated slopes. The initial value of
\(a=200\) was arbitrary and set after trying reasonable guesses that
converge to the slopes in the original article after 9 iterations for a
final value of \(a=159\).

\begin{Shaded}
\begin{Highlighting}[]
\NormalTok{a }\OtherTok{\textless{}{-}} \DecValTok{200}
\NormalTok{lypex\_ref }\OtherTok{\textless{}{-}} \FloatTok{1.058}
\NormalTok{tol }\OtherTok{\textless{}{-}} \FloatTok{0.001}
\NormalTok{lypex\_estimate }\OtherTok{\textless{}{-}} \DecValTok{2} \SpecialCharTok{*}\NormalTok{ lypex\_ref}
\NormalTok{iter }\OtherTok{\textless{}{-}} \DecValTok{0}

\ControlFlowTok{while}\NormalTok{ (}\FunctionTok{abs}\NormalTok{(lypex\_estimate }\SpecialCharTok{{-}}\NormalTok{ lypex\_ref) }\SpecialCharTok{\textgreater{}}\NormalTok{ tol) \{}
\NormalTok{  log\_of\_gravity}\SpecialCharTok{$}\NormalTok{log\_trade\_cens }\OtherTok{\textless{}{-}} \FunctionTok{log}\NormalTok{(a }\SpecialCharTok{+}\NormalTok{ log\_of\_gravity}\SpecialCharTok{$}\NormalTok{trade)}
\NormalTok{  log\_trade\_cens\_min }\OtherTok{\textless{}{-}} \FunctionTok{min}\NormalTok{(log\_of\_gravity}\SpecialCharTok{$}\NormalTok{log\_trade\_cens, }\AttributeTok{na.rm =} \ConstantTok{TRUE}\NormalTok{)}

\NormalTok{  fit\_tobit }\OtherTok{\textless{}{-}} \FunctionTok{censReg}\NormalTok{(}
    \AttributeTok{formula =} \FunctionTok{update.formula}\NormalTok{(ppml\_formula, log\_trade\_cens }\SpecialCharTok{\textasciitilde{}}\NormalTok{ .),}
    \AttributeTok{left =}\NormalTok{ log\_trade\_cens\_min,}
    \AttributeTok{right =} \ConstantTok{Inf}\NormalTok{,}
    \AttributeTok{data =}\NormalTok{ log\_of\_gravity,}
    \AttributeTok{start =} \FunctionTok{rep}\NormalTok{(}\DecValTok{0}\NormalTok{, }\DecValTok{2} \SpecialCharTok{+} \FunctionTok{length}\NormalTok{(}\FunctionTok{attr}\NormalTok{(}\FunctionTok{terms}\NormalTok{(ppml\_formula), }\StringTok{"term.labels"}\NormalTok{))),}
    \AttributeTok{method =} \StringTok{"BHHH"}
\NormalTok{  )}

\NormalTok{  lypex\_estimate }\OtherTok{\textless{}{-}} \FunctionTok{coef}\NormalTok{(fit\_tobit)[}\DecValTok{2}\NormalTok{]}
  \ControlFlowTok{if}\NormalTok{ (}\FunctionTok{abs}\NormalTok{(lypex\_estimate }\SpecialCharTok{{-}}\NormalTok{ lypex\_ref) }\SpecialCharTok{\textgreater{}} \DecValTok{2} \SpecialCharTok{*}\NormalTok{ tol) \{}
\NormalTok{    a }\OtherTok{\textless{}{-}}\NormalTok{ a }\SpecialCharTok{{-}} \DecValTok{5}
\NormalTok{  \} }\ControlFlowTok{else}\NormalTok{ \{}
\NormalTok{    a }\OtherTok{\textless{}{-}}\NormalTok{ a }\SpecialCharTok{{-}} \DecValTok{1}
\NormalTok{  \}}
\NormalTok{  iter }\OtherTok{\textless{}{-}}\NormalTok{ iter }\SpecialCharTok{+} \DecValTok{1}
\NormalTok{\}}
\end{Highlighting}
\end{Shaded}

\subsection{Non-Linear Least Squares}\label{non-linear-least-squares}

The starting values were retrieved from the PPML model results with zero
flows and then passed to the GLM function with a Gaussian log-link in R.

\begin{Shaded}
\begin{Highlighting}[]
\NormalTok{fit\_ppml\_eta }\OtherTok{\textless{}{-}}\NormalTok{ fit\_ppml\_2}\SpecialCharTok{$}\NormalTok{linear.predictors}
\NormalTok{fit\_ppml\_mu }\OtherTok{\textless{}{-}}\NormalTok{ fit\_ppml\_2}\SpecialCharTok{$}\NormalTok{fitted.values}
\NormalTok{fit\_ppml\_start }\OtherTok{\textless{}{-}}\NormalTok{ fit\_ppml\_2}\SpecialCharTok{$}\NormalTok{coefficients}

\NormalTok{fit\_nls }\OtherTok{\textless{}{-}} \FunctionTok{glm}\NormalTok{(}
\NormalTok{  ppml\_formula,}
  \AttributeTok{data =}\NormalTok{ log\_of\_gravity,}
  \AttributeTok{family =} \FunctionTok{gaussian}\NormalTok{(}\AttributeTok{link =} \StringTok{"log"}\NormalTok{),}
  \AttributeTok{etastart =}\NormalTok{ fit\_ppml\_eta,}
  \AttributeTok{mustart =}\NormalTok{ fit\_ppml\_mu,}
  \AttributeTok{start =}\NormalTok{ fit\_ppml\_start,}
  \AttributeTok{control =} \FunctionTok{list}\NormalTok{(}\AttributeTok{maxit =} \DecValTok{200}\NormalTok{, }\AttributeTok{trace =} \ConstantTok{FALSE}\NormalTok{)}
\NormalTok{)}
\end{Highlighting}
\end{Shaded}

\section{Replication results}\label{replication-results}

Santos Silva and Tenreyro (2006) is very close to full replication
according to the criteria defined in Peng (2011). The replication
results are consistent with the original article and reveal equivalent
figures for all the models estimated in the original article.

The results are presented in the following table:

\begin{Shaded}
\begin{Highlighting}[]
\FunctionTok{stargazer}\NormalTok{(}
\NormalTok{  fit\_ols\_1, fit\_ols\_2, fit\_tobit, fit\_nls, fit\_ppml\_1, fit\_ppml\_2,}
  \AttributeTok{header =} \ConstantTok{FALSE}\NormalTok{, }\AttributeTok{font.size =} \StringTok{"footnotesize"}\NormalTok{, }\AttributeTok{model.names =}\NormalTok{ F,}
  \AttributeTok{omit.table.layout =} \StringTok{"d"}\NormalTok{, }\AttributeTok{omit.stat =} \FunctionTok{c}\NormalTok{(}
    \StringTok{"f"}\NormalTok{, }\StringTok{"ser"}\NormalTok{, }\StringTok{"ll"}\NormalTok{, }\StringTok{"aic"}\NormalTok{, }\StringTok{"bic"}\NormalTok{, }\StringTok{"rsq"}\NormalTok{, }\StringTok{"adj.rsq"}
\NormalTok{  ),}
  \AttributeTok{title =} \StringTok{"Replication results for OLS (1{-}2), Tobit (3), NLS (4) and}
\StringTok{  PPML (5{-}6)."}
\NormalTok{)}
\end{Highlighting}
\end{Shaded}

\begin{table}[!htbp] \centering 
  \caption{Replication results for OLS (1-2), Tobit (3), NLS (4) and
  PPML (5-6).} 
  \label{} 
\footnotesize 
\begin{tabular}{@{\extracolsep{5pt}}lcccccc} 
\\[-1.8ex]\hline 
\hline \\[-1.8ex] 
 & \multicolumn{6}{c}{\textit{Dependent variable:}} \\ 
\cline{2-7} 
\\[-1.8ex] & (1) & (2) & (3) & (4) & (5) & (6)\\ 
\hline \\[-1.8ex] 
 lypex & 0.938$^{***}$ & 1.128$^{***}$ & 1.059$^{***}$ & 0.738$^{***}$ & 0.721$^{***}$ & 0.732$^{***}$ \\ 
  & (0.012) & (0.011) & (0.011) & (0.004) & (0.008) & (0.006) \\ 
  & & & & & & \\ 
 lypim & 0.798$^{***}$ & 0.866$^{***}$ & 0.848$^{***}$ & 0.862$^{***}$ & 0.732$^{***}$ & 0.741$^{***}$ \\ 
  & (0.011) & (0.011) & (0.010) & (0.005) & (0.008) & (0.006) \\ 
  & & & & & & \\ 
 lyex & 0.207$^{***}$ & 0.277$^{***}$ & 0.228$^{***}$ & 0.396$^{***}$ & 0.154$^{***}$ & 0.157$^{***}$ \\ 
  & (0.017) & (0.017) & (0.014) & (0.010) & (0.013) & (0.010) \\ 
  & & & & & & \\ 
 lyim & 0.106$^{***}$ & 0.217$^{***}$ & 0.178$^{***}$ & $-$0.033$^{***}$ & 0.133$^{***}$ & 0.135$^{***}$ \\ 
  & (0.017) & (0.017) & (0.014) & (0.007) & (0.013) & (0.010) \\ 
  & & & & & & \\ 
 ldist & $-$1.166$^{***}$ & $-$1.151$^{***}$ & $-$1.160$^{***}$ & $-$0.924$^{***}$ & $-$0.776$^{***}$ & $-$0.784$^{***}$ \\ 
  & (0.034) & (0.037) & (0.029) & (0.008) & (0.018) & (0.013) \\ 
  & & & & & & \\ 
 border & 0.314$^{**}$ & $-$0.241 & $-$0.225$^{**}$ & $-$0.081$^{***}$ & 0.202$^{***}$ & 0.193$^{***}$ \\ 
  & (0.143) & (0.164) & (0.109) & (0.010) & (0.034) & (0.026) \\ 
  & & & & & & \\ 
 comlang & 0.678$^{***}$ & 0.742$^{***}$ & 0.759$^{***}$ & 0.689$^{***}$ & 0.751$^{***}$ & 0.746$^{***}$ \\ 
  & (0.064) & (0.064) & (0.052) & (0.016) & (0.037) & (0.028) \\ 
  & & & & & & \\ 
 colony & 0.397$^{***}$ & 0.392$^{***}$ & 0.416$^{***}$ & 0.036$^{**}$ & 0.020 & 0.025 \\ 
  & (0.068) & (0.068) & (0.056) & (0.018) & (0.043) & (0.032) \\ 
  & & & & & & \\ 
 landl\_ex & $-$0.062 & 0.106$^{*}$ & $-$0.038 & $-$1.367$^{***}$ & $-$0.872$^{***}$ & $-$0.863$^{***}$ \\ 
  & (0.065) & (0.060) & (0.060) & (0.031) & (0.057) & (0.043) \\ 
  & & & & & & \\ 
 landl\_im & $-$0.665$^{***}$ & $-$0.278$^{***}$ & $-$0.478$^{***}$ & $-$0.471$^{***}$ & $-$0.703$^{***}$ & $-$0.696$^{***}$ \\ 
  & (0.063) & (0.060) & (0.059) & (0.022) & (0.054) & (0.040) \\ 
  & & & & & & \\ 
 lremot\_ex & 0.467$^{***}$ & 0.526$^{***}$ & 0.563$^{***}$ & 1.188$^{***}$ & 0.647$^{***}$ & 0.660$^{***}$ \\ 
  & (0.078) & (0.089) & (0.077) & (0.018) & (0.048) & (0.036) \\ 
  & & & & & & \\ 
 lremot\_im & $-$0.205$^{**}$ & $-$0.109 & $-$0.032 & 1.010$^{***}$ & 0.549$^{***}$ & 0.562$^{***}$ \\ 
  & (0.081) & (0.089) & (0.074) & (0.018) & (0.048) & (0.036) \\ 
  & & & & & & \\ 
 comfrt\_wto & 0.491$^{***}$ & 1.289$^{***}$ & 0.728$^{***}$ & 0.443$^{***}$ & 0.179$^{***}$ & 0.181$^{***}$ \\ 
  & (0.105) & (0.143) & (0.113) & (0.014) & (0.036) & (0.027) \\ 
  & & & & & & \\ 
 open\_wto & $-$0.170$^{***}$ & 0.739$^{***}$ & 0.310$^{***}$ & 0.928$^{***}$ & $-$0.139$^{***}$ & $-$0.107$^{***}$ \\ 
  & (0.049) & (0.048) & (0.040) & (0.024) & (0.039) & (0.029) \\ 
  & & & & & & \\ 
 logSigma &  &  & 0.677$^{***}$ &  &  &  \\ 
  &  &  & (0.007) &  &  &  \\ 
  & & & & & & \\ 
 Constant & $-$28.492$^{***}$ & $-$39.909$^{***}$ & $-$36.626$^{***}$ & $-$45.098$^{***}$ & $-$31.530$^{***}$ & $-$32.326$^{***}$ \\ 
  & (1.088) & (1.221) & (1.059) & (0.239) & (0.596) & (0.444) \\ 
  & & & & & & \\ 
\hline \\[-1.8ex] 
Observations & 9,613 & 18,360 & 18,360 & 18,360 & 9,613 & 18,360 \\ 
\hline 
\hline \\[-1.8ex] 
\textit{Note:}  & \multicolumn{6}{r}{$^{*}$p$<$0.1; $^{**}$p$<$0.05; $^{***}$p$<$0.01} \\ 
\end{tabular} 
\end{table}

\newpage

\section{Conclusion}\label{conclusion}

The replication effort was minimal, without the need to contact the
authors for clarifications or incur into data transformations or
filtering not mentioned in the article. The results are consistent with
the original article and reveal proper and transparent scholarship.
Ideally, all disciplines that rely on quantitative research and the use
of statistical methods to test hypotheses should follow similar or
higher standards of transparency, but that is not always the case as
initiatives such as Simonsohn, Nelson, and Simmons (2024) reveal.

\section{Acknowledgements}\label{acknowledgements}

Thanks to Joao Santos Silva for pointing that out, in a previous version
of this document I mentioned that the original results were obtained
with Stata.

\section*{References}\label{references}
\addcontentsline{toc}{section}{References}

\phantomsection\label{refs}


\end{document}